\input harvmac
\input epsf.tex
%\draftmode

\def \E   {{\rm E}}

\def \om {\omega}

\def \del {\partial}

\def \ha{{\textstyle{1\over 2}}}

\def \L {\Lambda}

\def \a {\alpha}
\def \b {\beta}

\def \H {\td H} 
\def\r {\rho}

\def \n {\nu}

\def \l {\lambda}

\def \td {\tilde }

\def \inv {^{-1}}
\def \ov {\over }

\def \lr { \lref}

\def \pl {{  Phys. Lett. }}

\def \prl {{  Phys. Rev. Lett. }}
\def \pr  {{ Phys. Rev. }}

\def \ijmp {{ Int. J. Mod. Phys. }}

\baselineskip8pt
\Title{
\vbox
{\baselineskip 6pt{\hbox{ }}{\hbox
{Imperial/TP/97-98/14}}{\hbox{hep-th/9712063}} {\hbox{
  }}} }
{\vbox{\centerline {Statistical mechanics of D0-branes }
\vskip4pt
 \centerline {and black hole  thermodynamics}
%\vskip4pt
% \centerline { }
}}
\vskip -27 true pt
\centerline  {  Hong  Liu{\footnote {$^*$} {e-mail address:
hong.liu@ic.ac.uk
 } } and 
 A.A. Tseytlin\footnote{$^{\star}$}{\baselineskip8pt
e-mail address: tseytlin@ic.ac.uk}\footnote{$^{\dagger}$}
{\baselineskip8pt  Also at  Lebedev  Physics
Institute, Moscow.} 
}
\bigskip
%\bigskip

%\centerline{{\it Theoretical Physics Group,  Blackett 
   %          Laboratory,  Imperial College,  London SW7 2BZ, U.K. }}
 \centerline {\it  Theoretical Physics Group, Blackett Laboratory,}
\smallskip
\centerline {\it  Imperial College,  London SW7 2BZ, U.K. }

\medskip\bigskip

\centerline {\bf Abstract}
\medskip
\baselineskip10pt
\noindent
We consider a system of D0-branes in toroidally compactified 
space with  interactions described by 
a Born-Infeld-type generalisation of the leading 
$v^2 + v^4/r^{D-4}$  terms  ($D$ is the number of non-compact 
 directions in M-theory, including the longitudinal one).
 This non-linear action  can be  interpreted 
as  an  all-loop  large $N$  super Yang-Mills effective action
and has a remarkable scaling property.
We first  study  the classical dynamics  of a brane probe 
in the field of a central brane source  and  observe  the interesting
difference between the $D=5$ and $D > 5$ cases:
 for $D >5$ the center acts as a completely absorbing
black hole of effective size 
proportional to a power of the probe  energy, while for $D=5$ there is no absorption for any impact parameter.
 A similar dependence on $D$
 is found  in 
the behaviour of the Boltzmann partition function  $Z$ of an ensemble of
 D0-branes.
 For $D=5$ (i.e. for compactification on 6-torus)  $Z$ is convergent at short distances and 
is analogous to the ideal gas one.  For $D > 5$ 
 the system  has short-distance instability.
For sufficiently low temperature  $Z$ is shown  to describe the thermodynamics
of a Schwarzschild black hole in $D > 5 $ dimensions, supporting 
recent discussions of black holes  in Matrix theory. 
\medskip
%%%%%%%%%%%%%%%%%%%%%%%%%%%%%%%%%%%%%%%%%%%%%%%%%%%%%%%%%
\Date {December 1997}
%%%%%%%%%%%%%%%%%%%%%%%%%%%%%%%%%%%%%%%%%%%%%%%%%%%%%%%%%%%%%%%%%%%
\noblackbox
\baselineskip 16pt plus 2pt minus 2pt

\def \gym {g_{\rm YM}} 
\def \rt {{\rm t}}

%%%%%%%%%%%%%%%%%%%%%%%%%%%%%%%%%%%%%%%%%%%%%%%%%%%%%
\lr \bfss{ T. Banks, W. Fischler, S. Shenker  and  L. Susskind, 
Phys. Rev. D55 (1997) 112,  hep-th/9610043. }

\lr \pol{J.  Polchinski, \prl 75 (1995) 4724.}

\lr \suss{ L. Susskind, hep-th/9704080.}
\lr\kraus{ E. Keski-Vakkuri and  P.  Kraus, 
%``Born-Infeld Actions from Matrix Theory",
 hep-th/9709122.}
     
\lr\bfks{ T. Banks, W. Fischler, I.R. Klebanov  and  L.  Susskind, 
hep-th/9709091.}
\lr\hm{ G.T. Horowitz  and  E.J. Martinec,
hep-th/9710217. }
\lr\li{ M. Li, hep-th/9710226.}
\lr\ks{ I.R. Klebanov and  L. Susskind, hep-th/9709108.}
\lr\gkp{
S.S.~Gubser, I.R.~Klebanov and A.W.~Peet, 
Phys. Rev. { D54} (1996) 3915, hep-th/9602135.}
\lr\kt{
I.R. Klebanov and A.A. Tseytlin, Nucl. Phys. { B475} (1996) 165,
hep-th/9604089.}
\lr\bfkss { T. Banks, W. Fischler, I.R. Klebanov and  L. Susskind, 
hep-th/9711005.}
\lr\bb{
K. Becker and M. Becker, 
%{\it A two loop test of M(atrix) theory},
hep-th/9705091.}
\lr\bbpt{
K. Becker, M. Becker, J. Polchinski and A.A. Tseytlin, 
%{\it Higher order graviton scattering in matrix theory},
\pr D56 (1997) 3174, 
hep-th/9706072.}
\lr\bek{ J. D. Bekenstein, Phys. Rev. D49 (1994) 1912.}
\lr\dm {S.R. Das, S.D. Mathur, S.K. Rama and  P. Ramadevi, 
hep-th/9711003.}
\lr\sei{N. Seiberg, ``Why is the Matrix Model Correct?" hep-th/9710009.}
\lr\sen{A. Sen, ``D0 Branes on $T^n$ and Matrix Theory" hep-th/9709220.}
\lr\dkps{ M.R. Douglas,
D. Kabat, P. Pouliot  and  S.H. Shenker, Nucl. Phys. B485 (1997) 85,
hep-th/9608024.}

\lr\land{L.D.  Landau and E.M. Lifschitz, ``Statistical Physics" (Pergamon, 1975). }
\lr\bac{C. Bachas, \pl B374 (1996) 37, hep-th/9511043.}

\lr \suus {L. Susskind, 
%``The world as a hologram",
 hep-th/9409089.}
\lr \hooft {
G. 't Hooft, gr-qc/9310026.}

\lr\dan{
U. Danielsson, G. Ferretti and B. Sundborg, \ijmp A11 (1996) 5463, hep-th/9603081;
D. Kabat and P. Pouliot, \prl 77 (1996) 1004, hep-th/9608024.}

\lr\chts{ I. Chepelev and   A.A. Tseytlin, hep-th/9709087. }
\lr\mald{J. Maldacena, hep-th/9709099.}

\lr\volovich{I. Volovich, hep-th/9608137.}
\lr\krauss{J. Harvey, hep-th/9706039;   P. Kraus, hep-th/9709199.}
\lr\dealwis{S. de Alwis,  hep-th/9710219.}
\lr\lars{V. Balasubramanian and F. Larsen, hep-th/9610077.}

\def \rx {{\rm x}}
\def\H {{\rm H}}
\def \V {{\cal V}}
\def \th {\theta}
\def \dth{\dot \theta}
\def \r {\dot r }
\def \r {\dot r }
\def \rb  {{\rm  b}}
\def \rr {{\rm r}}
\def \pp {{\rm p}}

\def \WW {{\cal W}}

%%%%%%%%%%%%%%%%%%%%%%%
\newsec{Introduction}
%%%%%%%%%%%%%%%%%%%%%%%%
In this paper we shall explore 
 some aspects of the classical mechanics and classical statistical 
mechanics  of an ensemble of D0-branes in toroidally compactified
string  theory
(with $d=D-1= 10-p$ non-compact dimensions), or, equivalently,  a
system of Dp-branes wrapped over the dual  torus. 
Our study is   motivated, in particular, 
 by   recent works on  black holes in Matrix theory
\refs{\bfks,\ks,\hm,\li,\bfkss,\dm}.

D-branes  are BPS  objects: if $N$  parallel branes of the same type
are located at some points in space the  interaction potential between them  vanishes \pol.  
%Put close to each other, they form a marginal bound state.
Given some kinetic energy, the branes are attracted  to each other
\refs{\bac,\dkps}.  One may thus  try to model a non-extremal 
D-brane with {\it large} charge $N$  by  a non-relativistic 
gas  of interacting 
 constituent unit-charge D-branes. In particular, one may try  to 
reproduce the entropy of the corresponding near-extremal Dp-brane
as  that of a  gas of  classical extremal Dp-branes. 
This picture is `dual' to   some  previous suggestions  \refs{\gkp,\kt}
of understanding the near-extremal Dp-brane  entropy in terms of 
quantum gases 
of   massless excitations  on brane world volumes.
The entropy of  classical near-extremal R-R p-branes 
can be expressed  in terms of the super Yang-Mills coupling constant 
$g_{\rm YM}$ of the world volume theory \ks\
\eqn\yyy{
S \ \sim \   [{E} ^{ {D-2 }     }\  V_{p}^{D-6 } \ N^{D-4} 
 (g^2_{\rm YM})^{{8-D}}]^{ 1 \ov 2 (D-4)}  \ 
,  \ \ \ \ \ \   \  D = 11-p
\ , } 
suggesting  a Yang-Mills interpretation \refs{\gkp,\kt,\bfks,\ks}. 

One may expect that  the two  (massive D0-brane gas and massless  YM gas) descriptions 
may be  
 useful  in the two different regions. Normally,  a 
massless gas will have a much higher entropy than a non-relativistic
gas  of  the same energy.
 Thus  when the  deviation $E$ of the energy from the extremal value 
 is large (but still small compared to the total energy of D-branes),
  the relevant  excitations
should  be mostly the  massless modes, i.e. 
the  massless gas picture 
should be  applicable. 
At the very low energy  the massless degrees of freedom 
in compact dimensions can no longer be excited, so that  
the description in terms of a non-relativistic gas of  D0-branes 
should be the adequate one. 

Reinterpreted in  the Matrix theory language (i.e. applied to   a Schwarzschild 
black hole in one higher dimension $D=d+1$
 with  a  compact light-like direction)
the description in terms of a gas of D0-branes
was recently discussed  in  \refs{\hm,\li,\bfkss}.
In the systematic picture   developed in \bfkss\
the D0-brane gas and the  SYM gas on the dual torus should be equivalent 
provided one takes into account  the effects due to off-diagonal gluons 
as short strings connecting D0-branes. As a result,  D0-branes  
in the gas should actually be   coupled to a an effective background  
 which makes them distinguishable \bfkss.

One may still hope to be able to describe some 
general  features of this unusual gas  by using mean field
approximation and  ignoring details 
of the background.
In  \bfkss\  it  was  suggested  to  view  the black hole  as  
a Boltzmann gas of distinguishable 
 D0-branes with  two-body interactions
given by the one-loop 1+0 SYM  effective action \refs{\dkps,\bfss}
\eqn\one{
L_{\rm eff} = \sum_{i} {v_{i}^{2} \over 2R} 
+ {c G_{D} \over R^{3}} \sum_{i,j} {v_{ij}^{4} \over r_{ij}^{n}}\ ,  }
where $n= D-4 =d-3= 7-p$. 
Applying qualitative considerations based on 
 the virial theorem, i.e.  setting  the 
first and the second term in \one\  to be  of the same order and 
assuming that $v \sim p/m \sim R/R_{s}$ (where $R$ is the radius of the light-like direction and $R_s$ is the size of D0-brane system),
one  obtains  \refs{\hm,\bfkss} the Schwarzschild black hole  expressions 
 for  the mass and the  entropy  of the system.

% and the correct interaction potential between the  two equal mass black holes
%(with $S \sim N$). 

Our point of departure is the following observation: 
at the virial  point where $ { G_{D} \over R^{2}} { v^2\ov r^{D-4}}  \sim 1$
the  higher-order $v^{2L+2}\ov r^{nL}$ terms  in the interaction potential
are all  of the same order and thus cannot be ignored a priori. 
Our  basic assumptions will be:
(i) large number $N$ of constituent branes and  (ii) small  and slowly changing  velocities (negligible accelerations). 
We shall also assume, in the spirit of mean field approximation, that 
under the conditions (i) and (ii) the 
forces of all other $N-1$  branes 
 acting on a given one  can be approximated 
by a   force exerted on a  unit charge   D-brane `probe' by a static 
 D-brane  `source'  of charge $N-1\approx N$  placed  at the center of the system.  
In our   crude model the   
 effect of the background  will be included only
in that the partition function of the classical 
D0-brane gas will not contain the usual $1/N!$ factor, 
reflecting the fact that constituent D0-branes should be treated as distinguishable.\foot{A 
suggestion that quantum extreme
 black holes described by  D-branes in the large $N$ limit   should obey Boltzmann infinite statistics was made in \volovich.}

There exists a natural candidate for the all-order 
generalisation of the $v^2 + {v^4\ov r^n}$ action \refs{\bbpt,\chts,\mald,\kraus,\dealwis}.  
Consider  the  classical  Born-Infeld
action  for a  D0-brane 
 probe moving  in a supergravity background
produced by a D0-brane  source  in $T^p$ compactified  space
(or T-dual configuration of a wrapped Dp-brane probe in 
Dp-brane source background on  $\td T^p$) 
\eqn\sug{
I= - m \int dt\ H_0\inv \big[\sqrt { 1 - H_0 v^2 } -1 \big] \ , \ }
\eqn\key{ H_0 = 1 + K \ ,  \ \ \  
\ \ \    K= { Q\ov r^n} \ , \ \ \ \ n=7-p=D-4 \ , \ \ \  r= |x_i| \ , \ \ \ 
v_i =\dot x_i \ , \ \ i=1,...,D-2 \ , } 
\eqn\qqu{ m = { 1 \ov g_s \sqrt{\a'}}=R\inv  \ , \ \  \ \
 Q= N g_s {(2\pi\sqrt {\a'})^7\ov V_p \omega_{6-p}}  \sim   N \gym^2 \sim
N G_D R^{-2}  \ ,  }
where
$V_p$ is the volume of the torus $T^p$ and 
 $ \gym^2 = (2\pi)^{p-2} (\sqrt {\a'})^{p-3} \td g_s $ is the YM coupling on 
the dual torus ($\td g_s = g_s (2\pi \sqrt{\a'})^p V_p\inv$). 
%and $a_p = 2^{2-p}\pi^{-(p+1)/2}{\Gamma ({\textstyle{ 7-p \ov 2}})}$. 
The supergravity action \sug\ is supposed to apply  only 
at {\it large} distances. 
 If we  take  the charge 
$N$ to be large (for fixed $r$) then  we can ignore the asymptotic value
  1 in the harmonic function,  replacing  $H_0$ 
 by $K$, i.e. replacing \sug\  by\foot{The same  action can be obtained by  a reduction  from 11 dimensions 
along the `null' \suss\ direction \refs{\bbpt,\chts} or by 
taking the low-energy (YM theory) limit \mald\
$\a'\to 0, \ \gym^2 \sim \td g_s (\sqrt {\a'})^{p-3} =$fixed, \
$ X_i = x_i/\a' =$fixed.} 
\eqn\sog{
S =  \int dt \ L  \ , \ \ \ 
L= - m  K\inv \big[\sqrt { 1 - K v^2 } -1 \big] 
= { m v^2 \ov 2} + { mQ v^4 \ov 8 r^n }  +{ mQ^2 v^6 \ov 16 r^{2n}  } + ...
 \  .    }
 This action can be interpreted as the supergravity action \sug\ 
formally 
extrapolated  to the short-distance (or near-horizon $r \approx 0$) region.  
It appears that this action captures some important
information about D0-brane dynamics at short distances.\foot{In general,  the quantum SYM effective  action (and its large $N$ Born-Infeld-type part)
contains  also fermionic contributions
 leading to spin-dependent terms in the interaction potential
\refs{\krauss,\li}. We  shall not include such terms 
in our model which is   based on mean field approximation.
We consider 
a field of a large number of branes  acting on a on a single brane in 
the gas so it is natural to treat 
interactions  as spherically symmetric
(spin effects should average to zero).}

It is {\it this} 
action  that  we shall use below, 
 assuming that  for {\it large} $N$ 
it  is the one that 
describes interactions of  slowly moving 
D0-branes  both at {\it short}  and {\it large} 
distances.  
Indeed, \sog\ can be interpreted  \refs{\chts,\mald} 
as  a sum of all   gauge-theory (light open string)
   loop corrections  to the 
low-energy interaction  potential 
between a pair of 
moving  Dp-branes, i.e. as the  large $N$ quantum effective action 
 of     $p+1$ dimensional super Yang-Mills  theory 
with the coupling $\gym$.  This identification
was explicitly checked at the 2-loop SYM level in \refs{\bb,\bbpt}. The miracle of maximal supersymmetry that relates 
the $v^4$ \dkps\ and $v^6$ \bbpt\ terms
 in the  long-distance  (classical supergravity)  and  short distance
(quantum  SYM)  string theory expression for the D-brane interaction 
potential 
 is conjectured  \refs{\chts,\mald}
 to continue (for  large $N$) to all orders in $v$.

Like its  $v^2 + {v^4\ov r^n}$ truncation, the full 
 action \sog\ has a remarkable  (`holographic' \refs{\dan,\dkps,\bfss})
  scaling property:
the dependence on the parameters $m$ and $Q$ ($\sim  N$) can be completely 
absorbed 
into redefinitions $x_i \to \l_1 x_i , \ t \to  \l_2 t$.
The resulting parameter-independent action  is  then  
\eqn\scal{
S = - \int d{\rm t}\   \rx^{n}\ \big[\sqrt { 1 - {\dot \rx^2 \ov \rx^n} } -1 \big] \ , }
\eqn\scall{
\rx_i =  m^{2 \ov n+2 } Q^ { - { 1 \ov n+ 2}}\  x_i  \ , \ \ \ 
\ \   {\rm t} = m^{n-2 \ov n+2 } Q^ { - { 2 \ov n+ 2}}\  t \ , \ \  
\ \ \ \ \dot \rx_i = {d\rx_i \ov d \rt} \ .  }

The  interaction   terms in the 
Lagrangian  in \sog\  are {\it attractive},  but 
the corresponding Hamiltonian  is  still positive
\eqn\ham{
\H =  p_i v_i - L =  m K\inv [\sqrt { 1 + K {  p^2 \ov m^2} } -1 ]  \ , 
  }
though this is not manifest if one  truncates  its 
expansion in powers of momenta, 
\eqn\eee{\H= {p^2 \ov 2 m} -   { Q p^4 \ov 4m^3r^n } + { Q^2 p^6 \ov 8m^5r^{2n} } +...\ , 
\ \ \ \ \ \ \ \ \  p_i = { mv_i \ov \sqrt { 1 - K  v^2}} \ . }
While the square root structure
of \sog\ and \ham\   is    reminiscent of the  relativistic 
particle expressions,  it should be  stressed again that \sog\
 applies only for sufficiently  small 
and slowly changing velocities  (note  that depending on $r$, 
the corresponding momenta may, however,  be large).

 Our aim   will be  to study  some consequences 
 of the   action \sog\ and apply it to a description 
of a statistical ensemble of D0-branes in order to clarify the relation
to  black hole thermodynamics.
% in \refs{\bfks, \hm,\li, \bfkss}.
The non-linear Born-Infeld-type actions  are known 
to have non-trivial classical dynamics with 
novel short-distance properties and  the present 
 case will not be an exception. Indeed, the structure of \sog,\scal\
suggest the presence of a dynamical bound $ {Q v^2\ov r^n } \leq 1$, which implies, 
in particular, that  
 $v$ must be small at small distances, in agreement with the 
non-relativistic nature of \sog. 
While at large distances  \sog\
is well-approximated by the $v^2 + {v^4\ov r^n}$ terms, 
  in the short distance region 
(and  the region of characteristic distances and velocities
satisfying ${ Q v^2\ov r^n } \sim 1$ or $p \gg 1$) 
  its predictions will be  very  different from 
 what would follow simply  from the $v^2 + {v^4\ov r^n}$  action.

We shall first consider the classical motion 
of a  0-brane probe in the field  of a  D0-brane source 
described by  the action \sog\ (Section 2).  We shall 
find  that the behaviour of the classical trajectories
has surprising dependence  on the number of compactified 
dimensions, i.e. on the power $n $ in $K$: 
the case of $n=1$ (corresponding  to D0-branes on $T^6$ 
 or to   wrapped 6-branes   on $\td T^6$, 
i.e. to  the  near-extremal black hole in  $d=4$
or the Schwarschild black hole in $D=5$) 
 is  quite  different from $n >  1$ ($D >5$) cases.\foot{The  special nature of the $n=1$ (and $n=2$) case
is suggested by the  fact that in the case of the radial motion
$v^2 = \dot \rx^2, $  \scal\ can be put into the form\ 
$
S= - \int d\rt  \ \mu (\rho)[  \sqrt { 1 - \dot \rho^2}-1 ] ,$\ 
where $ \ 
\mu  ={ m \rho^{-{2n \ov n-2}}
 \ , \ \ \rho = { 2 \ov n-2}   \rx^{ - { 1 \ov 2}(n-2)} }$
for $n > 2$ while for $n=1$  ($p=6$) \ 
$ \mu  = \ha m \rho^2 \ , \ \ \rho = \sqrt{ 2\rx }$
and for $n=2$ ($p=5$) \  \ $ \mu  = m e^{2 \rho} \ , \ \ \  \rho=\ln \rx .$}
For $n=1$ the probe with  a non-zero angular momentum  $J$ will 
always scatter away from the center, while for $n >1$ 
it  will always  fall  at the center if it  has a  sufficiently low
$J$ (but it takes an infinite time  for it to reach the core $r=0$).
For larger $J$ (or larger impact parameters)
the particle scatters away from the core.\foot{The classical scattering 
we discuss  can be viewed as a WKB approximation to  scattering of a 
quantum D0-brane on an extremal  D0-brane black hole in $D=10$, or as a scattering 
of a graviton (described by a curved space Laplace operator)
on a pp-wave (or infinitely boosted Schwarzschild) background 
in $D=11$.  Similar  dimension-dependent feature
($p=6$,  $p=5$  and $p < 4$ as special cases) 
of the  scattering of classical  minimally coupled scalar 
fields on Dp-brane backgrounds which is implied by  interplay between 
the attractive `Newtonian'  $1/r^{7-p}$ and repulsive centrifugal $1/r^2$
potential was  pointed out in \lars.}
This  suggest  a
 `black hole'  interpretation  of the attractive center and 
implies the existence  of  a  characteristic 
short-distance  scale  (the critical value of the impact parameter)  
$b \sim ({EQ\ov m})^{1\ov D-4}$.
A peculiar  feature of the action \sog\ 
(related to its $D=11$  fixed $p_-$ graviton eikonal scattering interpretation \bbpt) 
is that this  `horizon radius' 
depends on the energy $E$  of the  incoming particle.\foot{
It actually looks  
as if the incoming particle acts as a black hole with mass related to its kinetic energy.}
This translates into  the temperature dependence in the statistical mechanics 
context and is crucial for establishing the  relation 
to black hole thermodynamics.

In Section 3 we shall  consider the classical statistical 
mechanics of an ensemble of a large number $N$ 
of point particles  whose  interactions can be 
effectively described by a `mean field' Hamiltonian \ham.
Equivalently, the Boltzmann partition function   we shall compute
can be interpreted as describing a gas of non-interacting  classical 
D0 branes  in the external field  produced by a  central 
extremal D0-brane 
source (with the  `bootstrap' condition that the 
charge of the source is  chosen 
to be equal to the number of  particles $N$ in the surrounding 
cloud). 

Since the interaction potential in \sog,\ham\
vanishes at large distances, the  large-volume
 behaviour of the partition function $Z$ 
is  the same as for an ideal gas.  Remarkably, 
for $n=1$ ($d=4$ or $D=5$)  $Z$  is completely finite (in spite of an
 apparent singularity of the interaction part of  $\H$ at  $r\to 0$).
  In higher 
dimensions $n  > 1 $ ($d > 4$ or $D > 5$) the partition function 
is divergent at $r=0$.
 Surprisingly, this is analogous  to 
a similar dimension dependence observed  in the properties of  the classical 
dynamics in Section 2.\foot{Other arguments indicating that
 the cases  of $D=5$ and $D=6$ are special
were given  in \refs{\bfks,\ks,\bfkss}.}
 The  $n=1$ case does not admit a black hole interpretation, 
i.e. there is no phase transition and the system looks like an ideal gas 
even for small temperatures.  
We shall,  therefore, concentrate on the $n > 1$ case
  and suggest that the instability at small distances   is related 
to the  existence  of  the critical absorption  scale $b$ 
in the classical  scattering problem. 
We shall assume that the region of  sufficiently low temperatures when 
the  thermal length scale  $\l \sim \b^{-1/2}$ is of order of $b$
 is described by  the partition function where the integral over $x$
 is computed with a lower (short-distance) cutoff 
 of order $b$. The upper (volume) cutoff (taken  to be $\sim b + \l$) 
is then also of order $b$.

As a  result, the partition function becomes  a simple product of powers 
of $\b$ and $N$  and it  can be interpreted as that of a free gas of particles 
in  a $(D-2)$-dimensional {\it transverse}
 space (or `horizon sphere')  of size $R_s=b \sim (N/\b)^{1\ov D-4}$.
 Treating the particles as distinguishable ones  as suggested in \bfkss,  
 we will show in Section 4,  that 
the  corresponding  entropy is proportional to $N$.
Interpreting the energy of the system as the light-cone energy 
  \refs{\bfks}  we obtain  the same 
 entropy-mass relation as for a Schwarzschild black hole in $D$ dimensions, 
supporting  the previous arguments \refs{\hm,\bfkss}. 
%We shall also suggest an alternative interpretation 
%not based on matrix theory type interpretation
%in which this  system of  D0-branes describes 
%the thermodynamics of near-extremal D0-brane  with charge $N$.

%%%%%%%%%%%%%%%%%%%%%
\newsec{ Classical  dynamics}
%%%%%%%%%%%%%%%%%%%%%%
To gain some intuition about  dynamical  properties  of a system of D0-branes
which will be useful in the statistical mechanics context we shall 
first discuss the classical dynamics governed by the action  \sog.
Since this action describes  a single particle with mass $m$ 
moving in a velocity-dependent central force potential, 
the   angular momentum 
vector and the energy of the particle are conserved.
 As in 
the  standard central force problem, the particle motion will 
be planar.  The   Lagrangian  for the  coordinates in the plane of motion 
 $(r, \th)$ is  given by \sog\ with
 $$v^2 = \dot{r}^{2} + r^{2} \dth^{2}\ , $$
so that the conserved angular momentum  and the energy are 
\eqn\angul{
p_{\th} = {\del L \over \del  \dth} = {m r^{2} \dth \over \sqrt{1-k v^{2}}} =J\ , }
\eqn\en{
\H = p_i v_i - L = m K^{-1} ( {1 \over \sqrt{1-K v^{2}}} -1) = E\ 
\ . }
Solving for $v$ we get  
\eqn\vvv{
v^{2} = {2E \over m} { 1 + {E \over 2m} K  \over (1 + {E \over m} K)^{2} } \ ,  }
and thus 
\eqn\rrr{
\r^{2} =  {2E \over m} { 1 + {E Q\over 2mr^n} 
 \over (1 + {E Q\over mr^n } )^{2}} 
- {J^{2} \over m^{2} r^{2}} { 1 \over   (1 + {EQ \over mr^n} )^{2} }   \ . } 
%={2E \over m}  { r^n (r^n  + \ha b )  \over (r^n  + b)^{2}} 
%- {J^{2} \over m^{2} r^{2}} { r^{2n} \over   (r^n  + b)^{2}} }  
%\ , \ \ \ \ \    b \equiv {E \over m} Q \ . }
As  is clear from \en,\vvv, 
$ v^2$  and thus also $\r^2$  are  always bound by $K\inv = r^n/Q$, 
i.e. $ \r^2 < r^n/Q$. 
The dependence on  $Q$  and $m$  in all of the above relations
can be eliminated by  the scaling transformation
\scall\ (which  implies  also   a  rescaling  of $E$ but not of $J$).

To analyse the classical  trajectories  it is useful to 
 interpret \rrr\  in terms of a particle 
moving in  an effective   velocity-independent central  force 
potential $\V(r)$,  rewriting \rrr\  as
\eqn\efe{
\ha m \r^{2} + \V (r) = E \ , \ \ \ }
\eqn\pooo{
\ \
\V=  E\big[ 1 -  { 1 + {b^n\over 2r^n}  \over (1 + {b^n\over r^n } )^{2}}\big] 
+  {J^{2} \over 2 m r^{2}} { 1 \over   (1 + {b^n \over r^n} )^{2}}   \ ,  } 
where we have introduced the  characteristic  scale $b$, 
\eqn\scale{
b = ({E Q\over m})^{1\ov n} \ . }
At large distances $ r \gg b$  this  effective potential 
reduced to the sum of a 
{\it repulsive} $1\ov r^n$  potential\foot{The intuition 
that the large-distance potential should be  attractive  
which is based on  the  sign of the second term in the 
 expansion in \sog\
is misleading  since for a velocity-dependent potential one should 
first  fix the energy and then solve for velocity as  was done above (cf. also 
\eee).}
and the standard 
centrifugal term, 
\eqn\larr{
\V _{r\gg b } \to \  {3E^{2} Q\over 2 m r^n }  + {J^{2} \over 2 m r^{2}}
\ .  }
Since $\V \to 0$ for $r \to \infty$,  the  velocity  approaches
the constant asymptotic value $v_\infty = \sqrt { E \ov 2m }$.

At short distances $r\ll b$ \  the potential 
  $\V$ reduces  to a  sum of a {\it universal} (i.e. $E$-independent) {\it attractive}  $r^n$ term 
 and   the  repulsive centrifugal  term  suppressed by the $r^{2n}$, 
\eqn\short{
\V_{ r\ll b }\ \to \   E - {m \over 2 Q} r^{n} + {J^{2} \over 2 m b^{2n}  } r^{2(n-1)}
\ . } 
For $n> 2$ the centrifugal term is subleading at $r\to 0$, 
while for $n=2$  both terms are proportional to $r^2$. 
That means that for $n \geq 2$
(and sufficiently large $J$) 
 the potential  has  a 
maximum at some finite $r$, decreasing towards $E$ at $r=0$ and 
$0$ at $r\to \infty$. 
 For $n=1$ the centrifugal term  does not vanish at $r=0$
(it reduces to a constant
up to higher order $r^{3n}$ term) and the potential has a maximum at $r=0$.

The  details of  motion   of an incoming particle in  the  central force
 depend  on  relative values of $E$ and $J$ as well as on  the value 
of $n$.  To determine this dependence it  is useful to analyse  the possible 
turning points ($\r=0$) solving the equation \efe\ or $\V(r)=E$. The latter is  equivalent to 
\eqn\turn{
1 +  { b^n\over  2r^n }  =  { \rb^2 \over r^{2} }  \ , \  \ \ \ 
\ \ \ \  \ \ \ 
\rb \equiv  { J \ov m v_\infty} =
  { J \ov \sqrt { 2 m E}} \ , }
where $\rb$ is  the  impact parameter. 
 One special 
solution is $r=0$. 
 Since for $n > 1$  the derivative of the potential is zero at $r=0$, 
the particle  that can  reach
the center stops there. This is  not true
for $n=1$ where the force at $r=0$ is non-zero and  the particle
with $J=0$
reaches the center in finite time and bounces back.

For $n=1$   \turn\ becomes a quadratic equation  which always has a (positive) solution  for any non-vanishing $\rb$. That means that the particle with  $J\not=0$
is always scattered away, i.e. it can reach the center  
only   when  $J=0$.

For $n=2$  \turn\ reduces to $r^2 = \rb^2 - \ha b^2$ 
so that for large enough angular momentum  such that 
$\rb^2 > \ha b^2$  there is a turning point at finite $r$ while 
for  smaller  $J$ (for fixed  $E$) 
 corresponding to $\rb^2 <  \ha b^2$ the incoming  particle is 
eventually absorbed by the center. 
That means that the  effective  size of the absorbing center 
(corresponding to the critical case of $\rb^2 = \ha b^2$)
is  $\rb_* = { 1 \ov \sqrt 2}  { b }  =  \sqrt { EQ \ov 2m }$. 

For $n > 2$  one finds (first differentiating \turn\ to determine that the 
critical value of $r$ is equal to $ ({ n \ov 4} {b^n \ov \rb^2})^{ 1 \ov n-2}$) that  for 
\eqn\sss{
\rb\  > \ \rb_*\equiv  (2^{n-6} n^2)^{1 \ov 2n}\  b \   } 
 there are two  solutions $r_1,\ r_2$ of \turn\  
 (in addition to the 
$r=0$ one). The potential $\V$ is equal to $E$ at $r=0$, 
then decreases to a local minimum, then 
has a  maximum between $r_1$ and $r_2$ and  finally decreases 
to zero at large $r$ (see Figure).  

\vskip 0.5cm
\vbox{
\baselineskip=10pt
{\centerline{\epsfxsize=5.5in \epsfbox{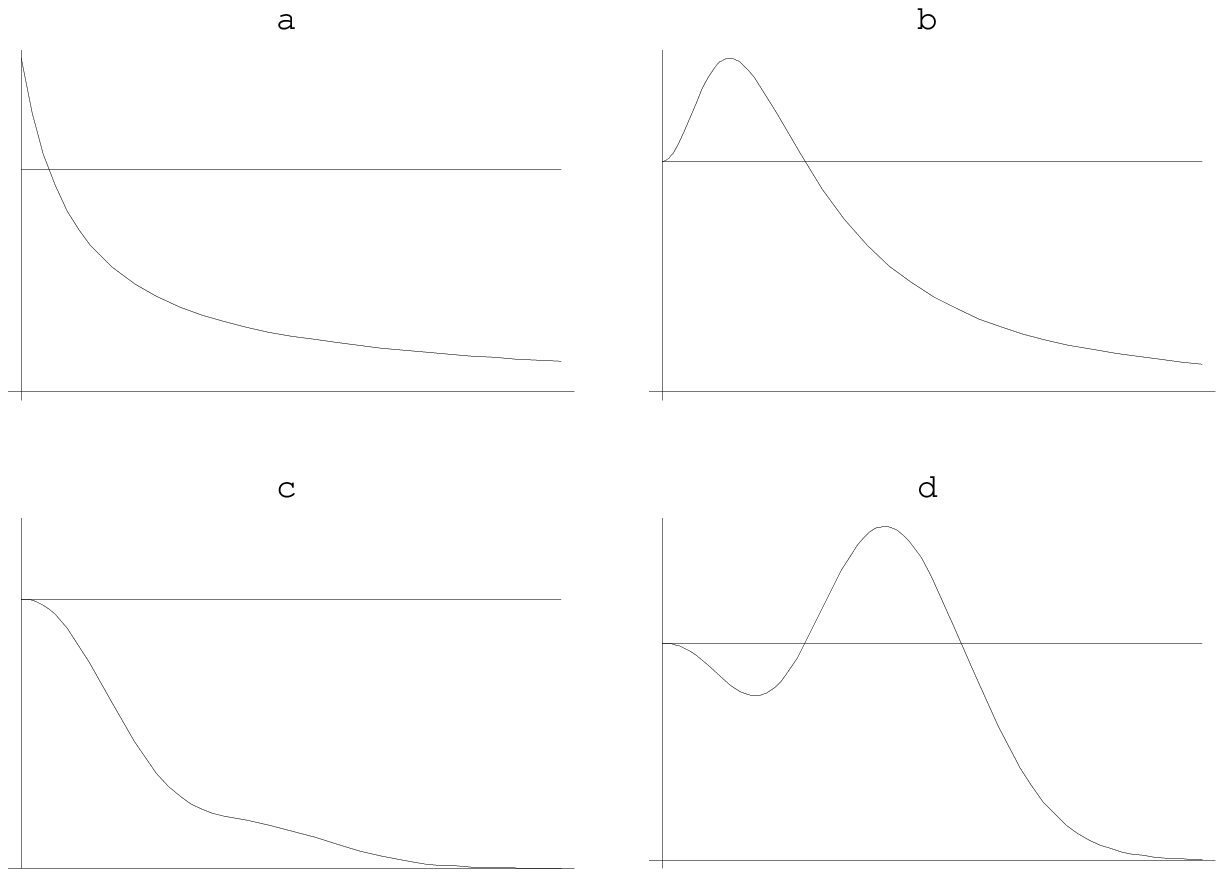}}}
\noindent{\bf FIGURE:}
\ninerm
Examples of effective potentials.
Plot (a) illustrates the form of  the effective potential $\V$ 
for  $n=1$ and $J \neq 0$, plot  (b) for $n=2$ and $\rb > \rb_*$,
(c) for $n > 1 $ and $\rb < \rb_*$, (d) for $n > 2 $ and $\rb > \rb_*$.
The upper horizontal lines on the plots
correspond to the  
energy of the particle $E$. 
}
\vskip .6cm

\baselineskip=16pt
\tenrm

The particle approaching the center  from infinity will  reach the minimal distance 
$r_2$ and then 
 scatter away.
A particle that at $t=0$ 
is  put  in the region $0 < r < r_1$  with some initial velocity 
directed out of the center will bounce once against the potential wall
at $r=r_1$ and then fall at the center.

For $\rb < \rb_*$ there are no turning points (the potential 
has its maximum  $E$ at $r=0$, see Figure)   and the incoming  particle is
always  absorbed
by the center.
The critical case $\rb=\rb_*$  defines\foot{The relation \sss\
is true  also for $n = 2$.}
 the  effective size of the 
 absorbing center (or `black hole' radius) 
 to be again of   order $b$, 
\eqn\size{
\rb_*  \sim  b \sim ({E\ov m} Q)^{ 1\ov n} \sim  (NE) ^{ 1\ov D-4} \ .
}
 This characteristic size of the system 
 is consistent with the  expectation (based on the square root form
of \sog,\en)  that  the 
critical  scale should be such that $Q {v^2\ov r^n} \sim 1 $, 
or, for $v=v_\infty = \sqrt{E\ov 2m}$,\ \  $r \sim b$. 
This is the same scale which appeared in the discussions
 \refs{\bfss,\hm,\bfkss} based on virial theorem considerations
applied to the leading-order $v^2 + { v^4\ov r^n} $ action. 
It can be interpreted as  the radius of a Schwarschild black hole
in one higher ($D=d+1$)  dimension (see Section 4).
% if $NE$ is identified with its energy. 
%In the context of the  `mean field' approximation 
%$NE$ is the kinetic energy of  constituents  of which the  black hole is built of.
Surprisingly, this interpretation does not apply for $n = 1$ ($D = 5$).

%%%%%%%%%%%%%%%%%%%%%
\newsec{ Statistical mechanics}
%%%%%%%%%%%%%%%%%%%%%%
In general, the  Boltzmann partition function of  a  system
of $N$ particles in $D-2=n+2$ spatial (`transverse') dimensions is 
\eqn\stat{
Z = { 1 \ov N!} \prod^{N}_{s=1}\int d^{D-2}p^{(s)}\  d^{D-2} x^{(s)} 
\, \exp \{ - \beta H_{tot} (p^{(1)},x^{(1)}, \cdots, p^{(N)},x^{(N)}) \} \ . }
As was already mentioned in the Introduction,  we shall 
 apply a
`mean field' approximation,  i.e.  make the assumption that the effect 
of interactions of 
each  constituent D0-brane  with all  others can be represented by the 
 effective
single-particle Hamiltonian \ham, i.e.
\eqn\haha{
H_{tot} (p^{(1)},x^{(1)}, \cdots, p^{(N)},x^{(N)})
= \sum_{s=1}^{N} \H (p^{(s)},x^{(s)}) \ , }
\eqn\hhh{
\H (p,x) = {m\ov Q}  r^{D-4} \big[  \sqrt { 1 +  
{Q \ov m^2}{ p^2 \ov  r^{D-4}}}  -1 \big] 
 \  , \ \ \ \  \ \ \ r= |x| \ .  }
We shall  follow   also the suggestion of 
\bfkss\ and  make the  drastic assumption that the  constituents
of this unusual  gas 
are to be treated  as distinguishable  so that one should
 drop the standard $ 1\ov N!$ factor in front
of $Z$. As we shall see, this is crucial for reproducing the standard 
Schwarschild black hole entropy-mass  relation.

The partition function  then  takes the form
\eqn\www{ 
Z =    W^{N} \ , \ \ \ \  \  \  \ \ 
W = \int d^{D-2}p \ d^{D-2} x  \, \  \exp\{- \beta \H( p,x)\} \ . } 
Notice that $Z$ depends on $N$ explicitly, as well as implicitly
via $Q$  \qqu\ in $\H$.

The integral over $p$ is  convergent  (it is  actually 
 the same  as in the case of a relativistic   ideal  gas \land\
  in  a general number of dimensions).\foot{Note that if one would  
formally keep only the $p^2$ and $p^4$ terms in 
the expansion \eee\ of $\H$ in \hhh\ the  integral over $p$ in 
 \www\  would be divergent.}
The integral over $x$ diverges at large  $r$ (since the potential part of 
$\H$ goes to zero at large distances) and, as for an ideal  gas, 
is to be cut off at the size    of the box
\eqn\sizz{
r < \L \ \sim \  V^{1 \ov D-2}\ , \  \ \ \ \ 
V=\om_{D-3} L^{D-2} \ . } 
 One finds  also  that  (for any value of $\b$) 
the  integral is 
{\it convergent }  at $ r\to 0$  if  $D=5$ ($n=1$)  but is {\it divergent}  
 if  $ D > 5$ ($n > 1)$. 

This is in  striking   analogy   with  the  properties 
of the classical motion  described by the action \sog\ or  Hamiltonian 
$\H$ \ham:
as we have seen in the previous section, for $n=1$ the  particle always scatters 
away from the center, while for $n > 1$ there exists a  special 
value of  the 
 impact parameter  $b$ \size\   below which the center acts as  a 
completely absorbing `black hole'.  
This  suggest   that for $n >1$ ($D > 5$) 
the system   is unstable and  to describe the   `black-hole phase'
one  should introduce
a short-distance cutoff  $  r > b$ where 
 $b$ \scale\   plays  
the role of  an intrinsic scale of the system. 
In the thermodynamical  context the energy of the  individual 
particle $E$ in \scale,\size\
should be  proportional to  the temperature,  $E \sim \b\inv$,  
so that  
\eqn\scl{
b(E)\  \to  \  b(\b)  = c_1 ({Q\over m\b})^{1 \ov D-4 } \sim ({N\ov \b})^{1 \ov D-4 } \ , }
where the coefficient  $c_1$ 
is a numerical 
 constant of order 1 (cf. \sss). 
Like for  a   Schwarzschild black hole, the
 characteristic scale thus depends on the temperature. 

In what follows we shall consider only 
the case of $D > 5$ which admits  a  black-hole 
interpretation. 
 We shall  further assume that 
for sufficiently low temperature
 corresponding to the black hole phase
the characteristic  size of the system 
or the  upper  cutoff $\L$ is  of the same order as 
the lower cutoff  $b$, i.e. 
\eqn\same{
 \L  = c_2  ({Q\over m\b})^{1 \ov D-4 } = \ k\    b  \  , \  \ \  \  \ 
 \ \  k={c_2 \ov c_1} > 1 \ , } 
so that  the particles are  distributed in a 
layer near the horizon, $  b <  r <  k  b$. 
This seems a natural  assumption 
since   one may expect that the 
difference  between the higher and lower cutoffs 
should be of the  order of  the thermal wave length $\l = \sqrt { \b/m}$, 
but, as we shall argue below,   at the  critical  temperature
$\l$  should be  of order  $b$. 

It is the  additional  dependence of  $Z$ on $\b$  and $N$  through 
the  cutoffs  $\L$ and $b$ 
 that makes the statistical mechanics of this system 
quite different from the usual  classical ideal gas one, 
but, at the same time, related to the black hole thermodynamics.

Integrating over the angles, one finds that   $W$ in \www\ is  given by 
\eqn\part{
W = (\om_{D-3})^2 \int^\L_b   dr\  r^{D-3} \int^\infty_0  dp\  
  p^{D-3}  \exp \{ - {\b m\ov Q} r^{D-4} [  \sqrt { 1 +  
{Q \ov m^2}{ p^2 \ov  r^{D-4}} }  -1 ] \}  . } 
Using the remarkable scaling symmetry of $D0$-brane system 
as described by \sog\ (cf.\scall), we can 
rescale  the coordinate and the momentum  
\eqn\resc{
 \rr = ({Q \over \beta m})^{ 1 \over D-4}\ r \ , \ \ \ \ \ \ \ 
\pp = ({m\over \beta })^{ 1 \over 2} p  \ , }
to  find the explicit dependence on the parameters
\eqn\prt{
W = (\om_{D-3})^2  
({Q \over \beta m})^{ D-2 \over D-4} 
({m\over \beta })^{ D-2 \over 2}\ 
    \WW (c_1, c_2)  \ , }
where 
\eqn\wwqw{
\WW = \int^{c_2}_{c_1}   d\rr\ \rr^{D-3} \int^\infty_0  d\pp \
  \pp^{D-3}\  \exp \{ - \rr^{D-4} \big[  \sqrt { 1 +  
{ \pp^2 \ov  \rr^{D-4}} } -1 \big] \}  \ .  } 
We have used the  crucial  observation   that  the rescaled values of the lower 
\scl\ 
and upper \same\  cutoffs 
are  simply numerical  constants 
\eqn\proo{ b \to    ({Q \over \beta m})^{ 1 \over D-4}  b  = c_1 \ ,\ \ \
  \ \ \ \ \ 
\L \to   ({Q \over \beta m})^{ 1 \over D-4}  \L  =c_2  \ .  }
Thus  $\WW  $  is  just   a   finite constant 
which does not depend on $\b$ and $N$.

Note that integrating over $\pp$  in \wwqw\ we get
\eqn\wwqq{
\WW = c_0  \int^{c_2^{  D-4}}_{c_1^{ D-4}}\   dy  \ 
  y^{D \over 2 (D-4)} \ e^y \  K_{D-1 \over 2} (y) \ , \ \ 
\ \ \ \ \ \ y\equiv \rr^{D-4}  \ , }
where $c_0$ is a numerical  constant and 
$K_\n$ is the  modified Bessel function.
Since  for $y \to 0$
 \  $  K_{\nu}(y) \to  y^{-\nu}
$  we have
$$y^{D \over 2 (D-4)} \ e^y \  K_{D-1 \over 2} (y)\  \to\ 
 y^{- {D^{2} +6D - 4 \over 2 (D-4)} } \ , 
$$
 so that, as  claimed above, 
  for $D >5$  the integral \wwqw\ 
 blows  up  in the limit  when the lower cutoff goes to zero ($c_1 \to 0$).

 %%%%%%%%%%%%%%%%%%%%%%%%%
\newsec{Relation to black hole  thermodynamics}
%%%%%%%%%%%%%%%%%%%%%%%%%%%%%%%
Eq.  \prt\ implies that the
  resulting partition function  \www\ is  thus  given by  
\eqn\zzzz{ 
  Z  = \  C \ [({Q \over \beta m})^{ D-2 \over D-4} 
({m\over \beta })^{ D-2 \over 2}]^N\  
   \sim \  \big({N \ov \beta^{D-2 \over 2} }\big)^{{D-2 \over D-4} N} 
\ , }
where $C$  is a   $\b, N $ -independent  numerical constant.
The  factor $  ({Q \over \beta m})^{ D-2 \over D-4} 
\sim  b^{D-2}\sim V $  is recognized (cf. \sizz--\same)   as the 
 $(D-2)$-dimensional  volume  
of the D0-brane gas, 
while  the second  
is the familiar factor of a power of
the mean thermal  wave length  
\eqn\ther{
\lambda = \sqrt { { \beta \over m}} \ . } 
The  partition function  \zzzz\ is thus the same   as 
that of a free gas  of ({\it distinguishable}) 
particles living on a $(D-2)$-dimensional  
transverse space of radius 
$R_{s} \sim b \sim ({Q \over m \beta })^{1\ov D-4} $ (which can be interpreted as  
the event horizon of a $D$-dimensional black hole),  i.e.\foot{
This interpretation in terms of a  gas    
on a $(D-2)$-dimensional  sphere may  be related to the  holographic 
principle \refs{\hooft,\suus}:
the partons should live in a space with  a non-trivial 
topology like a sphere. The restriction to the sphere
 may  be related to the presence of 
a  YM   background  discussed in \bfkss.}
\eqn\therm{
Z = \  C'\   ({V \over \lambda^{D-2}})^{N} \ . }
At the temperature so low that 
  the thermal wave length $\l$ 
of the  particles
 is of the order of the size of the box $b$, i.e.   $ \lambda^{D-2} \sim V$, 
one has 
$Z \sim 1$ and the   above expression for the  partition function 
 is no longer valid. It is natural to expect that  in this limit
the gas becomes  strongly correlated (degenerate)
and our classical approximation breaks  down.
 Note that  for  an
ideal gas this is the limit\foot{For the  usual ideal gas 
with $1/N!$ included in $Z$, the limit is actually 
$ \lambda^{D-2} \sim V/N$.}
 in which one  should 
 apply  quantum  statistics
(for a Bose gas, the corresponding  critical temperature is the one  
 at  which the Bose-Einstein condensation occurs).

We expect that the point $\l \sim b$  should 
correspond to a   phase transition  also in the  present case.
 This is  supported by the  resulting compelling 
interpretation in terms of   the $D$-dimensional 
 Schwarzschild black hole thermodynamics.
Indeed,  the limit  when 
\eqn\rea{ \l \sim b \ , \ \ \ \ \ \ {\rm i.e.} \ \ \ \  \ 
 ({Q \over \beta m})^{ D-2 \over D-4} \sim ({\beta \over m })^{ D-2 \over 2}
\ , }
corresponds to the critical temperature
\eqn\tema{
T = {\beta}\inv  \sim (G_{D}N)^{-{2 \over D-2}} R \ , }
and the size 
\eqn\sew{
R_{s} \sim b \sim  ({Q \over \beta m})^{1 \over D-4} \sim (G_{D} N)^{1 \over D-2}
\ , } 
where we used  \qqu.
The energy of the system 
 is  then given by
\eqn\fef{
\E = - {\del \ln Z \ov \del \b}  \sim \ N  T \   \sim G_{D}^{-{2 \over D-2}}  N^{D-4 \over D-2} R
\ , }
and the entropy  is found to be proportional to $N$, 
\eqn\rere{
S = \beta \E + \ln Z\  \sim\  N
\ , } 
where we have used that $\ln Z \ll N $ in the limit \rea. 

If we follow \refs{\bfss,\bfks} and interpret the energy $\E$
as 
the light-cone energy   related to the mass $M$  of a boosted object by 
\eqn\eqqq{
\E = { R \over N} M^{2} \ , }
then  using the expressions  \fef\  and  \rere\  derived above\foot{Eq.
\rere\ is equivalent to the condition $N \sim MR_s \sim S$ assumed in \bfks.}
we find  that 
\eqn\maass{
M \sim G_{D}^{-{ 1\over D-2}} S^{D-3 \over D-2}
 \ ,  \ \ \ { \rm i.e. } \ \ \ \ \ 
S \sim G_{D}^{1 \over D-3} M^{D-2 \over D-3}
 \ . }
This is 
precisely  the relation  characteristic to  a  Schwarzschild black hole
in $D$ dimensions.\foot{Note that
 keeping  the standard 
$ 1\ov N!$ factor (or $N^{-N}$ for large $N$) in  $Z$ \stat,\www,\zzzz\
would  give   the entropy-mass relation that is different from the 
Schwarzschild one.}

One obvious generalisation  of the discussion of this paper
is to replace the effective action \sog\
by that of a D0-brane probe moving in a background of {\it near-extremal}
D0-brane source. Indeed, it seems to be more consistent to treat the mean field
as that of a non-extremal charge $N$ D0-brane with excess energy 
being related to the kinetic energy of constituent  unit-charge D0-branes.
Another 
interesting problem  is to consider instead of  the classical partition function 
\stat,\www\  the partition function  of the quantum gas, i.e. the gas of quantum 
 D0-branes (or gravitons in $D=11$).  
It should be given   by the path integral 
of the exponent of the euclidean continuation of \sog\ (or its non-extremal version)
over periodic trajectories $x(\beta)=x(0)$ (the naive classical limit
$\beta \to 0$ gives back the classical $Z$). It may be that 
 the quantum partition function is  the proper starting point for trying to explain
the necessity to  introduce the short-distance cutoff  
(for $D > 5$) in the classical $Z$ and to drop the overall $1/N!$ factor.

 %%%%%%%%%%%%%%%%%%%%%%%%%
%\newsec{Concluding remarks}
%%%%%%%%%%%%%%%%%%%%%%%%%%%%%%%

%%%%%%%%%%%%%%%%%%%%%%%%%%%%%%%%
\newsec{Acknowledgements}
%%%%%%%%%%%%%%%%%%%%%%%%%%%%%
We are grateful to I. Klebanov for useful comments and explanations.
 We   acknowledge the support of PPARC and  the European
Commission TMR programme grant  ERBFMRX-CT96-0045.
%%%%%%%%%%%%%%%%%%%%%%%%%%%%%%%%%%%%%%%%%%%%%%%%%%%%%%%%%%%%%%%%%%%%%
\vfill\eject
\listrefs
\end